\documentclass[cits]{PoS}
\usepackage{graphicx}% Include figure files

\title{The $^{22}$Ne($\alpha$,n)$^{25}$Mg neutron source: latest experimental
results and prospects.}

\ShortTitle{$^{22}$Ne($\alpha$,n)$^{25}$Mg}

\author{\speaker{Claudio Ugalde}\\
        University of North Carolina at Chapel Hill\\
        E-mail: \email{cugalde@unc.edu}}

\abstract{The current status of the reaction rate of $^{22}$Ne($\alpha$,n)$^{25}$Mg is
summarized. Among the latest new results, probably the most relevant is 
the conclusion that the E$_x$=11.15 MeV 
state in $^{26}$Mg has a non-natural parity, so it does not contribute to the rates
of the $\alpha$ + $^{22}$Ne reactions. However, it may be possible that other 
neighboring states contribute to the neutron yield at stellar temperatures. 
Here we make an account of some of the experimental work in the literature that is
relevant to this state. Indeed, it would have been possible to avoid the controversy 
regarding this state before it even started.}

\FullConference{10th Symposium on Nuclei in the Cosmos\\
		 July 27 - August 1 2008\\
		 Mackinac Island, Michigan, USA}

\begin{document}

\section{Introduction}
The $^{22}$Ne($\alpha$,n)$^{25}$Mg reaction is considered to be the main
neutron source for the s process in core He-burning massive stars and 
also is of relevance in He-shell burning in AGB stars. By influencing 
the abundance of low mass s-process nuclei, this reaction also affects  
the seeds for the p process. For example, some of these nuclei, like 
the light Mo and Ru isotopes, are underproduced unless the rate for
the $^{22}$Ne($\alpha$,n)$^{25}$Mg reaction is somehow enhanced\cite{Costa:2000}. 
 
The $^{22}$Ne($\alpha$,n)$^{25}$Mg reaction competes with 
the $^{22}$Ne($\alpha$,$\gamma$)$^{26}$Mg process for the available 
$^{22}$Ne. These nuclei are produced by the 
$^{14}$N($\alpha$,$\gamma$)$^{18}$F($\beta$)$^{18}$O($\alpha$,$\gamma$)$^{22}$Ne 
chain of reactions in a stellar environment rich both in $^{4}$He and $^{14}$N
nuclei left from hydrogen burning via the CNO cycle. The temperature at which the 
neutron production is activated is such that the ratio of the two reaction rates 
is close to unity\cite{Karakas:2006}. Therefore, the efficiency of this neutron source is 
regulated by this ratio, so both reaction rates need to be determined 
simultaneously.

Inside the Gamow peak in these stellar scenarios, the $^{22}$Ne+$\alpha$ processes
are mostly resonant and involve the formation of the $^{26}$Mg 
compound nucleus. Direct measurements of the cross
section are challenging due to the Coulomb barrier, so currently available
reaction rates require an extrapolation to the lowest energies. Also, a guess
of the cross section requires information of the structure of $^{26}$Mg at excitation
energies in the Gamow peak, such as excited states, their energies, spins, and parities.
On the other hand, partial widths (or spectroscopic factors) of the different channels 
involved in the process need to be known as well. 

Here, a summary of what is known about the $^{22}$Ne($\alpha$,n)$^{25}$Mg reaction
at temperatures relevant to nucleosynthesis in the s process will be given.

%\begin{figure}
%\includegraphics[width=6.6cm]{levelsMg26.eps}
%\includegraphics[width=9.6cm]{measuredCross.eps}
%\caption{\label{fig:levelsMg26}(Left) Level scheme (not drawn to scale)
%of $^{26}$Mg showing the $^{22}$Ne + $\alpha$ entrance channel and the
%two competing exit channels $^{25}$Mg + n and $^{26}$Mg + $\gamma$. 
%Note the negative Q-value 
%(Q = -478.3$\pm$0.04 keV) for the $^{22}$Ne($\alpha$,n)$^{25}$Mg process.
%(Right) Yield curve for the $^{22}$Ne($\alpha$,n)$^{25}$Mg reaction 
%assembled from Jaeger et al. and Drotleff et al's results. The vertical solid black 
%lines correspond to the Gamow peak limits for T$_9$ $\leq$ 0.4. The horizontal 
%colored lines show the sensitivity limits for three direct measurements.}
%\end{figure}

\section{The reaction rate}
Several direct measurements have been performed to determine 
the reaction rates at energies that may have relevance in stellar scenarios 
(for example, see  
\cite{Ashery:1969,Haas:1973,Wolke:1989,Harms:1991,Drotleff:1991,
Drotleff:1993,Giesen:1993,Jaeger:2001}). So far, the best 
sensitivity has been achieved by Jaeger et al.\cite{Jaeger:2001} at 
$\sigma \sim$ 10$^{-11}$ barn. Experiments have reached 
energies inside the Gamow window, but still, the extrapolation of the cross
section to lower energies introduces important uncertainties
to the rate estimate.

It has long been thought that a resonance at E$_{lab}$=635 keV dominates 
the stellar reaction rates. Berman et al.\cite{Berman:1969} first proposed 
its existence based on their observation of a state in  $^{26}$Mg 
at E$_x$=11.15 MeV. They claimed the state to 
have J$^{\pi}$=1$^{-}$ based on a couple of arguments. First, a comparison 
of their 90$^o$ and 135$^o$ photo-neutron cross section measurements
favored a $\pi$=- assignment over $\pi$=+ \footnote{This fact is only 
mentioned in their paper but the cross section at 90$^o$ is not 
shown, so the actual comparison can not be assessed by the reader.}. 
Second, they compared electron inelastic scattering measurements
by Titze and Spamer\cite{Titze:1966} and Bendel et al. \cite{Bendel:1968} 
at two angles and various energies. Their analysis of the distribution 
of the strengths between magnetic and electric transitions suggested 
J$^{\pi}$=1$^{-}$, a result that contradicted Bendel et al.'s own 
conclusion the previous year. 

It is likely that the first indication of 
the M1 nature of the ground state transition for the E$_x$=11.15 MeV 
state in $^{26}$Mg comes from the parallel works of Titze 
and Spamer, and Bendel et al. However, the resolution of their experiments 
was such that it is not possible to establish a clear 1 to 1 correspondence 
between their state and that from Berman et al.'s high 
resolution work.

The confirmation of the spin-parity nature of the E$_x$=11.15 MeV 
state came years later when Crawley et al.\cite{Crawley:1989} performed 
inelastic proton scattering experiments on $^{26}$Mg at 201 MeV. 
They measured angular 
distributions in the range 2.5$^o$ $\leq$ $\theta_{c.m.}$ $\leq$ 12.5$^o$
and based on distorted wave Born approximation (DWBA) and distorted wave 
impulse approximation (DWIA) analyses, were able to identify 19 states 
with J$^{\pi}$=1$^{+}$. The state at E$_x$=11.15 MeV was among them.

Soon after Crawley et al.'s result was published, experimental efforts for 
measuring this resonance directly via the $^{22}$Ne($\alpha$,n)$^{25}$Mg
reaction were presented. For example, Harms et al.\cite{Harms:1991} 
found some evidence of the existence of the resonance by measuring the 
neutron yield at one single beam energy (E$_{\alpha}$=635 keV). Drotleff 
et al.\cite{Drotleff:1991} also reported a resonant structure found with their 
windowless gas target system at E$_{\alpha}$=623 keV, but later concluded 
it to belong to the background $^{11}$B($\alpha$,n)$^{14}$N reaction \cite{Drotleff:1993}.
Since both $^{22}$Ne and $\alpha$-particles have a ground state 
with J$^{\pi}$=0$^{+}$, only natural parity states in  $^{26}$Mg can be 
populated via $^{22}$Ne + $\alpha$ reactions. Therefore, the
$^{22}$Ne($\alpha$,n)$^{25}$Mg and $^{22}$Ne($\alpha$,$\gamma$)$^{26}$Mg 
reactions can not show the resonant structure at E$_{lab}$=635 keV.
However, this state could be of importance in the competing neutron poison
reaction $^{25}$Mg(n,$\gamma$)$^{26}$Mg.

Data reanalyses and compilations have also been published since then.
The works of the NACRE collaboration\cite{Angulo:1999}, K\"appeler et 
al.\cite{Kappeler:1994}, Koehler\cite{Koehler:2002}, and Karakas et 
al.\cite{Karakas:2006} have all computed the reaction rates assuming 
the  E$_x$=11.15 MeV state in $^{26}$Mg to have a natural parity, 
following Berman et al.'s\cite{Berman:1969} suggestion. The most recent 
direct measurement available (Jaeger et al.) and the indirect measurements 
below the neutron threshold of Ugalde et al.\cite{Ugalde:2007} provide 
a calculation of the reaction rate for the ($\alpha$,n) and ($\alpha$,$\gamma$)
processes, respectively, under the same assumption.

There are other recent experiments that support the non-natural parity 
assignment of Crawley et al. 
For example, Tamii et al.\cite{Tamii:2007} developed a technique to measure 
proton inelastic scattering angular distributions at forward angles with 
high resolution using the Grand Raiden spectrometer at Osaka. For $^{26}$Mg(p,p')$^{26}$Mg 
their resolution was 17 keV, good enough to identify the 
E$_x$=11.15 MeV state in $^{26}$Mg and then assign to it 
a J$^{\pi}$=1$^{+}$\cite{Fujita:2008}. Tonchev et al.\cite{Tonchev:2008} performed 
a $^{26}$Mg($\gamma$,$\gamma$)$^{26}$Mg experiment with the Free Electron Laser 
facility at Duke University. Their polarized $\gamma$-ray beam impinged on a 
$^{26}$MgO target and the outgoing photons were observed both at parallel and 
perpendicular positions with respect to the beam polarization plane with Ge detectors. 
They determined the transition from the E$_x$=11.15 MeV state to the ground 
state of $^{26}$Mg to be of an M1 character. Finally, Ugalde et al.\cite{Ugalde:2008} 
searched for the  E$_x$=11.15 MeV state in $^{26}$Mg with the $^{22}$Ne($^6$Li,d)$^{26}$Mg 
$\alpha$-particle transfer reaction and obtained a negative result, concluding that 
either the state has non-natural parity or the $\alpha$-particle spectroscopic factor
is very small. Either way, this state does not contribute to the rates for the 
$^{22}$Ne + $\alpha$ reactions. 

\begin{figure}
\begin{center}
\includegraphics[width=9.6cm]{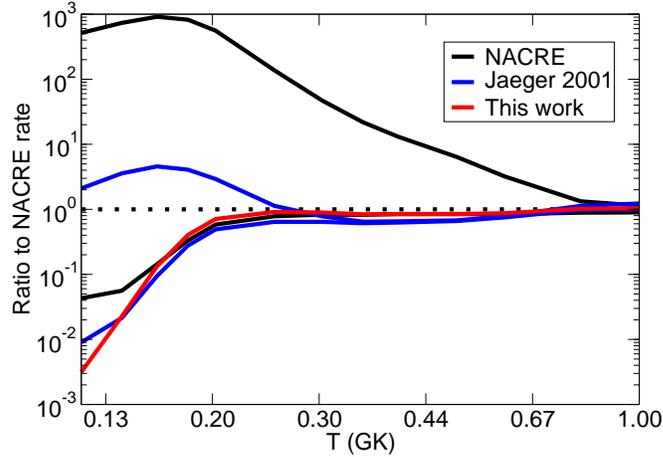}
\caption{\label{fig:rate_ne22an}
(Color) Rate for the $^{22}$Ne($\alpha$,n)$^{25}$Mg reaction normalized to the values adopted 
by the NACRE collaboration\cite{Angulo:1999}. For comparison, the rate values computed in the  
direct measurement of Jaeger et al. are shown as well. Upper and lower limits of the new rate 
are presented and discussed in \cite{Ugalde:2008}.    
}
\end{center}
\end{figure}

A calculation of the reaction rate 
for $^{22}$Ne($\alpha$,n)$^{25}$Mg based on Crawley et al.'s conclusion 
is shown in figure \ref{fig:rate_ne22an}. For example, this rate is in good agreement 
with the lower values suggested by the NACRE collaboration. However, NACRE's upper value can 
be rejected \cite{Ugalde:2008}. One of the main consequences of this result is that, as discussed
by Costa et al. \cite{Costa:2000}, the production of the light isotopes of Ru and Mo 
may require also a contribution from nucleosynthesis in accreting neutron stars or black holes.      

There are other states around the E$_x$=11.15 MeV state that may 
contribute to the ($\alpha$,n) and ($\alpha$,$\gamma$) reaction rates. Based on new experimental
results, this possibility will be discussed by Ugalde et al.\cite{Ugalde:2008}.

\section{Conclusion}
We have discussed some of the latest experimental results relevant to 
the reaction rate of $^{22}$Ne($\alpha$,n)$^{25}$Mg. In particular, 
it has been shown how it is possible to resolve the controversy regarding 
role of the state at E$_x$=11.15 MeV in $^{26}$Mg to the rate. This is interesting 
as experimental work conclusive enough has existed almost as long as the
controversy itself. It is important to stress too, that a lot of work remains 
to be done, specially with the $^{22}$Ne($\alpha$,$\gamma$)$^{26}$Mg reaction.

\bibliographystyle{plain}
\bibliography{ugalde_nicX}% Produces the bibliography via BibTeX.

\end{document}